\documentclass[useAMS,usenatbib,usegraphicx,referee]{mn2e}

\title[Intergalactic absorption]
{A Monte Carlo simulation of the intergalactic absorption and the
detectability of the Lyman continuum from distant galaxies}
\author[A. K. Inoue and I. Iwata]
{Akio K. Inoue$^{1}$\thanks{E-mail: akinoue@las.osaka-sandai.ac.jp
(AKI)} and Ikuru Iwata$^{2}$\\
$^{1}$College of General Education, Osaka Sangyo University, 3-1-1,
Nakagaito, Daito, Osaka 574-8530, Japan\\
$^{2}$Okayama Astrophysical Observatory, National Astronomical
Observatory of Japan, Kamogata, Okayama 719-0232, Japan}
\begin{document}

\date{Accepted Received; in original form}

\pagerange{\pageref{firstpage}--\pageref{lastpage}} \pubyear{2007}

\maketitle

\label{firstpage}

\begin{abstract}
We have made a Monte Carlo simulation of the intergalactic absorption in
 order to model the Lyman continuum absorption, which is required to
 estimate the escape fraction of the Lyman continuum from distant
 galaxies. To input into the simulation, we derive an empirical
 distribution function of the intergalactic absorbers which reproduces
 recent observational statistics of the Lyman $\alpha$ forest, Lyman
 limit systems (LLSs), and damped Lyman $\alpha$ systems (DLAs)
 simultaneously. In particular, we assume a common functional form of
 the number evolution along the redshift for all types of absorbers. 
 The Lyman series transmissions in our simulation reproduce the observed
 redshift evolution of the transmissions excellently, and the Lyman
 continuum transmission also agrees with an observed estimation which is
 still quite rare in the literature. The probability distribution of the
 Lyman $\alpha$ opacity in our simulation is log-normal with a tail
 towards a large opacity. This tail is produced by DLAs. The probability
 distribution of the Lyman continuum opacity in our simulation also show
 a broad tail towards a large opacity. This tail is produced by
 LLSs. Because of the rarity of LLSs, we have a chance to have a clean
 line of sight in the Lyman continuum even for $z\sim4$ with a
 probability of about 20\%. Our simulation expects a good correlation
 between the Lyman continuum opacity and the Lyman $\alpha$ opacity,
 which may be useful to estimate the former from the latter for an
 individual line of sight. 
\end{abstract}


\section{Introduction}

Neutral hydrogen remaining in the intergalactic medium (IGM) 
absorbs the radiation from distant sources \citep[e.g.,][]{gun65}. 
This intergalactic absorption seems to be caused by numerous discrete
systems, which are called absorbers in this paper, on the line of
sight. The absorbers probably consist of various types of objects: the
outer edge of galaxies, halo gas, diffuse medium in the intergalactic
space, etc. They probably connect with each other and form the 
``cosmic web'' \citep[e.g.,][]{rau98}.

Because of such clumpy nature of the IGM, the intergalactic absorption
fluctuates among the lines of sight. Thus, we cannot predict
the amount of the absorption for a certain line of sight. However, 
we can predict a mean amount of the absorption with its standard
deviation for many lines of sight in a statistical sense 
\citep[e.g.,][]{mol90,gia90,zuo93,mad95}. \cite{ber99} showed that a
Monte Carlo simulation is very powerful to discuss not only the mean
absorption but also the dispersion correctly. Here we present such a
simulation. Because many observational results of the IGM absorbers have
been published since 1999, we update the absorbers' statistics to be
input into the simulation.

Recently, \cite{mei06} presented a new mean curve of the intergalactic
absorption based on the density distribution produced by a cosmological
simulation. However, he discussed mean transmission only. 
\cite{tep07} have done a Monte Carlo simulation and discuss the
distribution of the absorption amount. However, they concentrate on only
Lyman $\alpha$ absorption. 
In this paper, we will examine the stochastic nature not only of the
Lyman series absorption but also of the Lyman continuum one. 

This is motivated by the recent observational attempts for determining
the escape fraction of the Lyman continuum, especially the hydrogen
ionising photons, from distant galaxies 
\citep[e.g.,][]{ste01,gia02,mal03,ino05,sha06,sia07}. 
In the process towards the escape fraction, we need a correction of
the intergalactic absorption for the Lyman continuum of the distant
galaxies. Since the Lyman continuum absorption by the IGM has not been
measured well observationally, a model predicting the absorption amount
of the Lyman continuum is required.

The effective optical depth through a clumpy IGM 
for the frequency $\nu_{\rm S}$ in the rest-frame of a source at the
redshift $z_{\rm S}$ is \citep[e.g.,][]{par80}  
\begin{equation}
 \tau_{\rm eff}(\nu_{\rm S}, z_{\rm S}) = \int_0^{z_{\rm S}} dz 
  \int_{N_{\rm l}}^{N_{\rm u}} d N_{\rm HI} 
  \frac{\partial^2 {\cal N}}{\partial z \partial N_{\rm HI}} 
  (1-e^{-\tau_{\rm cl}})\,,
\end{equation}
where $\partial^2 {\cal N}/\partial z \partial N_{\rm HI}$ is the number
of absorbers along the line of sight per unit redshift $z$
interval and per unit HI column density $N_{\rm HI}$ interval, and 
$\tau_{\rm cl}=\sigma_{\rm HI}(\nu_{\rm S}(1+z)/(1+z_{\rm S}))N_{\rm HI}$ 
is the optical depth of an absorber with $N_{\rm HI}$ at $z$,  
with being the HI cross section $\sigma_{\rm HI}(\nu)$ at the frequency
$\nu$ in the absorber's rest-frame. 
If we assume the column density distribution of the absorbers as a
power-law with an index $-\beta$ independent of the redshift, 
which is well found in observations \citep[e.g.,][]{tyt87}, we have 
\begin{equation}
 N_{\rm HI} \frac{\partial \tau_{\rm eff}}{\partial N_{\rm HI}} 
  \propto N_{\rm HI}^{1-\beta} (1-e^{-\tau_{\rm cl}})
  \propto \cases{
  N_{\rm HI}^{2-\beta} & ($\tau_{\rm cl}\ll1$) \cr
  N_{\rm HI}^{1-\beta} & ($\tau_{\rm cl}\gg1$) 
  }\,.
\end{equation}
Since observations show $\beta\approx1.5$ \citep[e.g.,][]{tyt87}, 
the maximum contribution to $\tau_{\rm eff}$ is made by absorbers with 
$\tau_{\rm cl}\sim1$ \citep[e.g.,][]{mol90,mei06}. Thus, the absorption
of the Lyman continuum is mainly caused by the Lyman limit systems
(LLSs) with $N_{\rm HI}\sim10^{17}$ cm$^{-2}$, not by the Lyman $\alpha$
forest (LAF) with $N_{\rm HI}\sim10^{13}$ cm$^{-3}$. Therefore, 
we take a great care of the treatment of LLSs in order to properly 
compute the Lyman continuum absorption by the IGM. Moreover, the Lyman
continuum absorption should be very stochastic because LLSs are
relatively rare. Therefore, we also need to model the dispersion of the
absorption correctly.

Previous works about the intergalactic absorption assumed different
number evolutions along the redshift for the LAF and LLSs. \cite{mad95}
assumed $d{\cal N}/dz \propto (1+z)^{2.46}$ for the LAF \citep{mur86} 
and $d{\cal N}/dz \propto (1+z)^{0.68}$ for LLSs \citep{sar89}. 
\cite{ber99} assumed a different set of the LAF number evolution and 
the same LLS evolution as \cite{mad95} \citep[see also][]{tep07}.
\cite{mei06} assumed the number evolution extracted from
a cosmological simulation for the LAF and  
$d{\cal N}/dz \propto (1+z)^{1.5}$ for LLSs \citep{ste95}.
However, \cite{ino05} showed in their appendix that recent LLSs
observations reported by \cite{per03} can be reproduced by a common
number evolution function for both types of absorbers. We take this
common function scenario in this paper because of its simplicity.

The rest of this paper consists of 5 sections; in section 2, we present
an empirical distribution function of the intergalactic absorbers and
show that the function reproduces all the observational data
simultaneously. In section 3, we describe the procedure of our Monte
Carlo simulation. In section 4, we describe the method for calculating
the IGM transmission, compare our transmission model with others, show
that our model reproduces the observed Lyman series transmissions very
well, predict the Lyman continuum transmission, discuss the distribution
functions of the intergalactic opacities of the Lyman continuum as well
as Lyman $\alpha$ line, and present a method for estimating the Lyman
continuum opacity from the Lyman $\alpha$ one. In section 5, we discuss
the detectability of the Lyman continuum from distant galaxies with our
Monte Carlo results. The final section is devoted to our conclusions.

\section{Distribution function of intergalactic absorbers}

In this section, we present an empirical distribution function of the
intergalactic absorbers which is input into the Monte Carlo simulation
in the later sections. We consider a functional form with several
parameters which are determined so as to reproduce the observed
redshift, column density, and Doppler parameter distributions of the
intergalactic absorbers. Especially, we assume a common number evolution
function along the redshift for all types of absorbers. 

An absorber have three physical quantities characterising itself:
redshift $z$, HI column density $N_{\rm HI}$, and Doppler parameter
$b$. Here, we assume that these three quantities are independent of each
other. This is just for simplicity although a slight correlation between 
$N_{\rm HI}$ and $b$ has found \citep[e.g.,][]{kim01}. Then, we express
the number of absorbers per unit volume of the three parameters' space as 
\begin{equation}
 \frac{\partial^3 {\cal N}}{\partial z \partial N_{\rm HI} \partial b} 
  = f(z) g(N_{\rm HI}) h(b)\,.
\end{equation}
Note that we have assumed a common functional form of the number
evolution along the redshift for all types of absorbers as $f(z)$. 
In the following, we determine the functional forms of $f(z)$, 
$g(N_{\rm HI})$, and $h(b)$ so as to reproduce the observed distribution
functions. Since the observed statistics of absorbers are not very tight
yet, we do it by visual inspection instead of doing it by a rigorous
way like a likelihood method.

\begin{figure}
 \includegraphics[width=8cm]{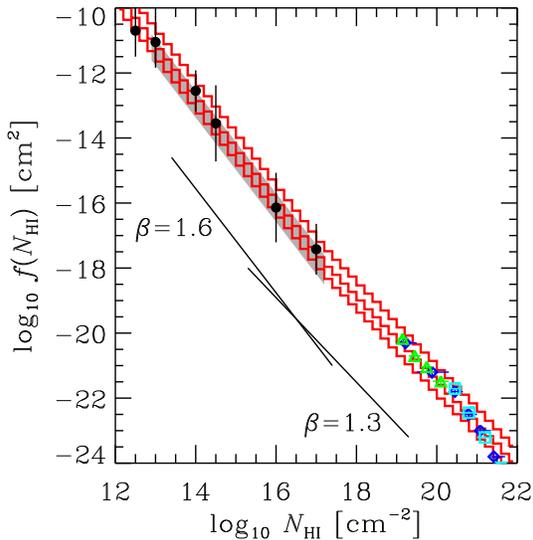}
 \caption{Number of the intergalactic absorbers per unit column
 density per unit ``absorption distance'' along an average line of sight
 as a function of the column density. The shaded area is the observed
 range of the Lyman $\alpha$ forest (LAF) at $z_{\rm abs}=0.5$--1.9 by
 Janknecht et al.~(2006). The filled circles are the data of the LAF at 
 $z_{\rm abs}=1.5$--4.0 based on Kim et al.~(2002). The diamonds,
 triangles, and squares are the observed data of sub-damped Lyman
 $\alpha$ systems (sub-DLAs) and damped Lyman $\alpha$ systems (DLAs) by 
 Peroux et al.~(2005; $z_{\rm abs}=1.8$--5.0), O'Meara et al.~(2007; 
 $z_{\rm abs}=1.8$--4.2), and Prochaska et al.~(2005; 
 $z_{\rm abs}=2.2$--5.5), respectively. All the data are scaled for a
 flat $\Lambda$ cosmology with $\Omega_\Lambda=0.7$.
 The histograms are the column density distributions of absorbers
 generated by our Monte Carlo simulation: the absorber's redshift 
 $z_{\rm abs}=0$--2 (bottom), 2--4 (middle), and 4--6 (top). The
 vertical shifts of the histograms are the result of the number density
 evolution along the redshift. The two solid lines are for the reference
 of the power-law index.}
\end{figure}

\begin{figure}
 \includegraphics[width=8cm]{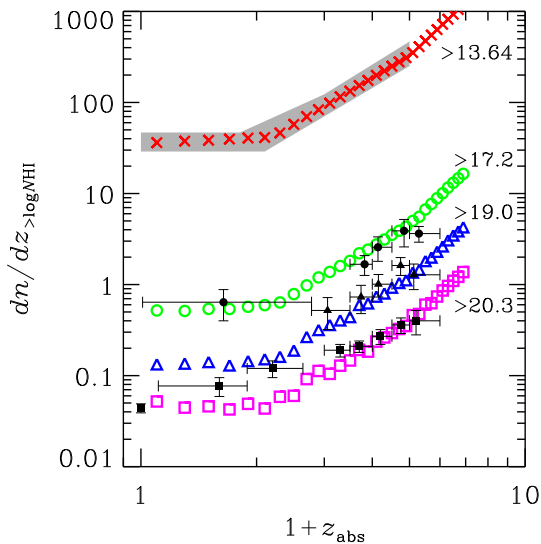}
 \caption{Number of the intergalactic absorbers per unit
 redshift along an average line of sight as a function of the absorbers'
 redshift. The shaded area is the observed range for absorbers with 
 $\log_{10} (N_{\rm HI}/{\rm cm}^{-2})>13.6$ (LAF) taken from Weymann et
 al.~(1998), Kim et al.~(2001), and Janknecht et al.~(2006). The filled
 circles, triangles, and squares are the observed data of absorbers with
 $\log_{10} (N_{\rm HI}/{\rm cm}^{-2})>17.2$ (LLS) taken from Peroux et
 al.~(2005), $>19.0$ (sub-DLA) taken from Peroux et al.~(2005), and 
 $>20.3$ (DLA) taken from Rao et al.~(2006), respectively. The number
 density evolutions of absorbers generated by our Monte Carlo simulation
 are shown by different symbols depending on the column density range:
 crosses ($\log_{10} (N_{\rm HI}/{\rm cm}^{-2})>13.6$), 
 open circles ($\log_{10} (N_{\rm HI}/{\rm cm}^{-2})>17.2$), 
 open triangles ($\log_{10} (N_{\rm HI}/{\rm cm}^{-2})>19.0$), and 
 open squares ($\log_{10} (N_{\rm HI}/{\rm cm}^{-2})>20.3$). They
 excellently trace the assumed distribution functions of absorbers given
 in equations (3)--(6), but we see statistical fluctuations in large
 column density cases due to the small number of such absorbers.}
\end{figure}

Figure~1 shows the column density distribution. The vertical axis is the
number of absorbers per unit column density and per unit ``absorption
distance'' which was introduced to remove the effect of the Hubble
expansion \citep{bah69}. While \cite{tyt87} proposed a single power-law
for the column density distribution, the recent data seems a double
power-law with a break at $\sim10^{17}$ cm$^{-2}$ as found in Figure~1. 
\cite{pro05} also suggest a similar break around $10^{17}$ cm$^{-2}$. 
For the highest column density, however, there may be a more rapid
decline of the number of absorbers, suggesting a Schechter like function
for this column density range \citep{per03,pro05}. Here we adopt a
double power-law for simplicity as 
\begin{equation}
 g(N_{\rm HI}) = B \cases{
  \left(\frac{N_{\rm HI}}{N_{\rm c}}\right)^{-\beta_1} 
  & ($N_{\rm l} \leq N_{\rm HI} < N_{\rm c}$) \cr
  \left(\frac{N_{\rm HI}}{N_{\rm c}}\right)^{-\beta_2} 
  & ($N_{\rm c} \leq N_{\rm HI} \leq N_{\rm u}$)
 }\,, 
\end{equation}
with $\beta_1=1.6$ and $\beta_2=1.3$ obtained by eye. 
The break column density $N_{\rm c}$ is assumed to be 
$1.6\times10^{17}$ cm$^{-2}$ which is the usual criterion for LLSs,
i.e. the absorber with a column density larger than this value is
optically thick for the Lyman limit photon. A physical explanation for
this break, which gives more number of higher column density absorbers,
is a self-shielding effect; the neutral fraction inside an optically
thick absorber (i.e. LLSs) could be kept higher. The normalisation $B$
is determined by 
$\displaystyle \int_{\rm N_{\rm l}}^{N_{\rm u}} g(N_{\rm HI}) dN_{\rm HI}=1$.

Figure~2 shows the number evolution along the redshift of the LAF, LLSs,
sub-DLAs, and DLAs. The LAF number evolution (the shaded area) shows a
break at $z\sim1$ \citep{wey98}. \cite{dav99} have explained that this
break is due to a steep decline of the ionising background intensity
from $z\sim1$ to 0. Thus, we assume a break at $z\sim1$ in the
functional form of $f(z)$. On the other hand, \cite{fan06} suggest a
steepening of the Lyman $\alpha$ opacity evolution at $z\ga4$, which
means more absorbers than expected at high redshift. As done in
\cite{ino06}, thus, we assume another break at $z\sim4$ in
$f(z)$. Therefore, we assume  
\begin{equation}
 f(z) = {\cal A} \cases{
  \left(\frac{1+z}{1+z_1}\right)^{\gamma_1} & ($0 < z \leq z_1$) \cr
  \left(\frac{1+z}{1+z_1}\right)^{\gamma_2} & ($z_1 < z \leq z_2$) \cr
  \left(\frac{1+z_2}{1+z_1}\right)^{\gamma_2}
  \left(\frac{1+z}{1+z_2}\right)^{\gamma_3} & ($z_2 < z$)
  }\,.
\end{equation}
We adopt $z_1=1.2$, $\gamma_1=0.2$, and $\gamma_2=2.5$ which are
determined by eye to fit the observed number evolutions along the
redshift of the LAF in Figure~2. We also adopt the second break in
$f(z)$ at $z_2=4.0$ with $\gamma_3=4.0$ to reproduce a rapid increase of
the Lyman $\alpha$ opacity found by \cite{fan06}. With these parameters,
the observed number evolutions of all types of absorbers are reproduced
simultaneously. Note that we have assumed $g(N_{\rm HI})$ to be a double
power-law as equation (4). If we assumed $g(N_{\rm HI})$ to be a single
power-law, we could not reproduce the number evolution functions of all
types of absorbers simultaneously.  The normalisation $\cal A$
is the total number of the IGM absorbers at $z=z_1$ with a column density 
$N_{\rm l} \leq N_{\rm HI} \leq N_{\rm u}$. We adopt ${\cal A}=400$ 
with $N_{\rm l}=10^{12}$ cm$^{-2}$ and $N_{\rm u}=10^{22}$ cm$^{-2}$ 
to match with the observed number of the LAF in Figure~2.

For the Doppler parameter distribution function $h(b)$, we assume the
functional form with a single parameter $b_\sigma$ suggested by
\cite{hui99}:
\begin{equation}
 h(b) = \frac{4 {b_\sigma}^4}{b^5} e^{-{b_\sigma}^4/b^4} \,.
\end{equation}
We adopt $b_\sigma=23$ km s$^{-1}$ based on the
measurements by \cite{jan06}. We note that $h(b)$ in equation (6) is
already normalised as $\displaystyle \int_0^\infty h(b) db = 1$.

In Figures~1 and 2, we show comparisons of the empirical distribution
functions presented in equations (4) and (5) with the observed
functions. For our distribution functions, we show the results generated
by our Monte Carlo simulation, whose procedures are described in the
next section. We have confirmed that our Monte Carlo simulation
reproduces the input distribution functions excellently. 
Figure~1 shows that our column density distribution $g(N_{\rm HI})$ is
very consistent with the observations. Note that the vertical axis in
the panel is not exactly same as $g(N_{\rm HI})$ but the column density
distribution averaged over a redshift range (and divided by the
``absorption distance''). Figure~2 shows that our redshift evolution
$f(z)$ nicely reproduces the observed redshift evolution for all types
of absorbers simultaneously. Note that the vertical axis in the panel is
not exactly same as $f(z)$ but the number density of absorbers with
column densities larger than a certain value. Although the disagreement
of DLAs at $z\sim1$ (but less than a factor of two) might suggest a
redshift dependence of $g(N_{\rm HI})$, we avoid this complexity. The
adopted values for the parameters are summarised in Table~1. 

\begin{table}
 \caption[]{Parameters for the distribution function of intergalactic
 absorbers.}
 \setlength{\tabcolsep}{3pt}
 \footnotesize
 \begin{minipage}{\linewidth}
  \begin{tabular}{lc}
   \hline
   Parameters & Adopted value\\
   \hline
   $\gamma_1$ & 0.2\\
   $\gamma_2$ & 2.5\\
   $\gamma_3$ & 4.0\\
   $z_1$ & 1.2\\
   $z_2$ & 4.0\\
   ${\cal A}$ & 400\\
   $\log_{10} (N_{\rm l}/{\rm cm}^{-2})$ & 12.0\\
   $\log_{10} (N_{\rm u}/{\rm cm}^{-2})$ & 22.0\\
   $\log_{10} (N_{\rm c}/{\rm cm}^{-2})$ & 17.2\\
   $\beta_1$ & 1.6\\
   $\beta_2$ & 1.3\\
   $b_\sigma/{\rm km~s}^{-1}$ & 23.0 \\
   \hline
  \end{tabular}
 \end{minipage}
\end{table}%

Figure~3 is the close-up of the number evolution of LLSs along the
redshift. As described in section 1, we need to reproduce the observed
number of LLSs for a rigid prediction of the Lyman continuum
absorption by the IGM. Figure~3 ensures a good agreement between our
model and observations by \cite{per05}. On the other hand, 
the regression line suggested by \cite{sar89} shown as the dotted line
in Figure~3, which is assumed in \cite{mad95}, \cite{ber99}, and
\cite{tep07}, does not agree well with the recent data. The dashed line
in Figure~3 is the regression line suggested by \cite{ste95} which is
assumed in \cite{mei06}. This case agrees with the recent data. However,
in this case, we would need different functional forms of $f(z)$ for the
LAF and for LLSs to fit all the data shown in Figure~2 simultaneously.

\begin{figure}
 \includegraphics[width=8cm]{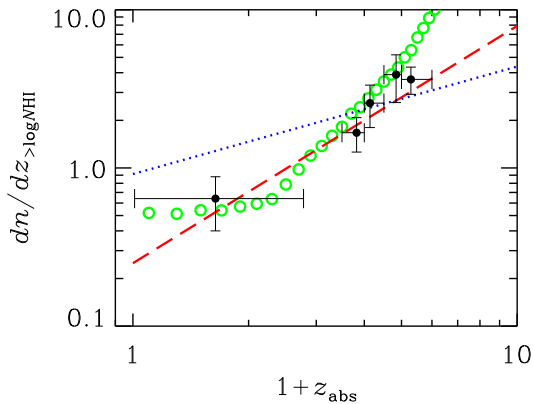}
 \caption{Number of LLSs per unit redshift along an average line of
 sight as a function of the LLSs' redshift. The filled circles are the
 recent observed data by Peroux et al.~(2005). The open circles are our
 model. The dashed line is the regression line by Stengler-Larrea et
 al.~(1995), which is assumed in Meiksin (2006). The dotted line is the
 regression line by Sargent et al.~(1989), which is assumed in Madau
 (1995), Bershady et al.~(1999), and Tepper-Garc{\'i}a \& Fritze~(2008).}
\end{figure}

\section{Monte Carlo procedure}

Based on the assumed distribution function described in the previous
section, we generate a large number of absorbers along lines of sight 
by a Monte Carlo method. The random number generator used 
in this paper is the ``Mersenne Twister'' developed by \cite{mat98} 
which is a very fast random number generator with a very good
statistical property.
\footnote{The ``Mersenne Twister'' is proved to have the period of 
$2^{19937}-1$ and the 623-dimension equidistribution property. Its
C-code is public at  
http://www.math.sci.hiroshima-u.ac.jp/\~{}m-mat/MT/emt.html}

We assume the encounter of an absorber on a line of sight to be 
a Poisson process. In other words, we neglect the effect of 
clustering of absorbers although it has been already found
\citep{ost88}. We should examine the effect in future.
For a Poisson process, if we have an absorber at $z$, 
the probability encountering the next absorber at $z+\Delta z$ is 
\begin{equation}
 p(\Delta z; z) = f(z)e^{-f(z) \Delta z}\,,
\end{equation}
because the mean redshift interval of two absorbers at $z$ is 
just the reciprocal of the redshift distribution function $f(z)$ 
of equation (5).

For a line of sight, we start from $z=0$ and determine 
the redshift of the first absorber by drawing a random number 
based on equation (7), and we determine the column density 
and the Doppler parameter of this absorber by drawing other 
two random numbers based on $g(N_{\rm HI})$ and $h(b)$ 
of equations (4) and (6). 
We note that $g(N_{\rm HI})$ and $h(b)$ themselves are 
the probability distribution functions. Then, we determine 
the redshift, the column density, and the Doppler parameter 
of the next absorber likewise. These procedures are repeated 
until the absorber's redshift exceeds $z=6$. Typically, 
we generate about 18,000 absorbers for one line of sight which 
depends on the lower limit of the column density $N_{\rm l}$. 
For a smaller $N_{\rm l}$, we need much more absorbers which 
is time-consuming. As shown in equation~(2), absorbers with 
a column density less than $N_{\rm l}=10^{12}$ cm$^{-2}$ do not
contribute to the opacity very much because their optical depths are
much less than unity. 
We generate 10,000 lines of sight for enough statistics.

\section{Intergalactic transmissions}

Based on the intergalactic absorbers along lines of sight generated 
by the Monte Carlo procedures described in the previous section, 
we calculate transmissions along the lines of sight.

Suppose an absorber with the redshift $z$, the column density 
$N_{\rm HI}$, and the Doppler parameter $b$. The optical depth 
of this absorber for the frequency $\nu$ in the absorber's 
rest-frame is 
\begin{equation}
 \tau(\nu) = N_{\rm HI} \left(\sigma_{\rm LC}(\nu) 
  + \sum_i \sigma_i(\nu) \right)\,,
\end{equation}
where $\sigma_{\rm LC}$ is the cross section 
for the Lyman continuum, and $\sigma_i$ is the cross section 
of the $i$-th Lyman series line. The cross section 
for the Lyman continuum is approximated to 
\begin{equation}
 \sigma_{\rm LC}(\nu) = \sigma_{\rm LL} 
  \left(\frac{\nu_{\rm LL}}{\nu}\right)^3 
  \hspace{1cm}(\nu \geq \nu_{\rm LL}) \,,
\end{equation}
where the cross section at the Lyman limit $\nu_{\rm LL}$ is 
$\sigma_{\rm LL}=6.30\times10^{-18}$ cm$^2$ \citep{ost89}. When 
$\nu < \nu_{\rm LL}$, $\sigma_{\rm LC}(\nu) = 0$.
The cross sections for the Lyman series lines are approximated to 
\begin{equation}
 \sigma_i(\nu) = \frac{\sqrt{\pi} e^2 f_i}{m_{\rm e} c \nu_{\rm D}} 
  \phi_i(\nu)\,,
\end{equation}
where $m_{\rm e}$ is the electron mass, $e$ is the electron charge, 
$c$ is the speed of light, $f_i$ is the oscillator strength of the 
$i$-th line, $\nu_{\rm D} = \nu_i (b/c)$ is the Doppler width 
with the line centre frequency $\nu_i$, and $\phi_i(\nu)$ is the 
line profile function which is calculated by an analytic approximation
by \cite{tep06}.
The hydrogen atomic data of $\nu_i$, $f_i$, and the damping constant
$\Gamma_i$ for the line profile up to 40-th line are taken from
\cite{wie66}.

Since we consider only H{\sc i} absorbers in this paper, 
we calculate the wavelength range not affected by helium:  
700--1300 \AA\ in the rest-frame of the source redshift.
Note that we should consider wavelength enough longer than 
that of the Lyman $\alpha$ line (1216 \AA) in order to treat a broad 
wing of damped Lyman $\alpha$ lines. We should also calculate 
transmissions with an enough high wavelength resolution. 
We have found experimentally that typical errors in transmissions within
a wavelength range relative to those calculated with 0.01 \AA\
resolution in the source rest-frame are 10\% for 1 \AA\ resolution and
1\% for 0.1 \AA\ resolution.\footnote{We should note that relative
errors are larger for lower transmissions. For a very small
transmission, say $\la10^{-3}$, the relative error becomes about 10\% or
more in some cases even with 0.1 \AA\ resolution. Such cases appear in
the Lyman continuum transmissions for the source redshift $z_{\rm S}\ga5$. 
However, the fraction of the cases is $\sim10\%$.}
Thus, we have calculated all the results in this paper with
0.1 \AA\ resolution in the source rest-frame.

\subsection{Example and average transmissions}

Figure~4 shows the IGM transmissions along an example 
line of sight. We find a lot of narrow absorption lines as the LAF. 
In the Lyman continuum region, we find several step-wise depressions 
caused by LLSs. For example, in the panel (b) ($z_{\rm S}=3$ case), 
we find significant transmissions below the Lyman limit up to 864 \AA\ 
at which a sharp depression appears because of a LLS at $z=2.78$ which
has $N_{\rm HI}=6.5\times10^{17}$ cm$^{-2}$. This means that the
Lyman continuum absorption by the IGM is very stochastic because it
is controlled by the presence of relatively rare LLSs. This point is
illustrated in Figure~5. If we do not have a LLS (or DLA) near the
source, we can expect a significant transmission even far below the
source Lyman limit (panel [a]). However, If we have such a LLS, the
transmission is suddenly cut down at the corresponding wavelength
(panel [b]).

\begin{figure*}
 \includegraphics[width=15cm]{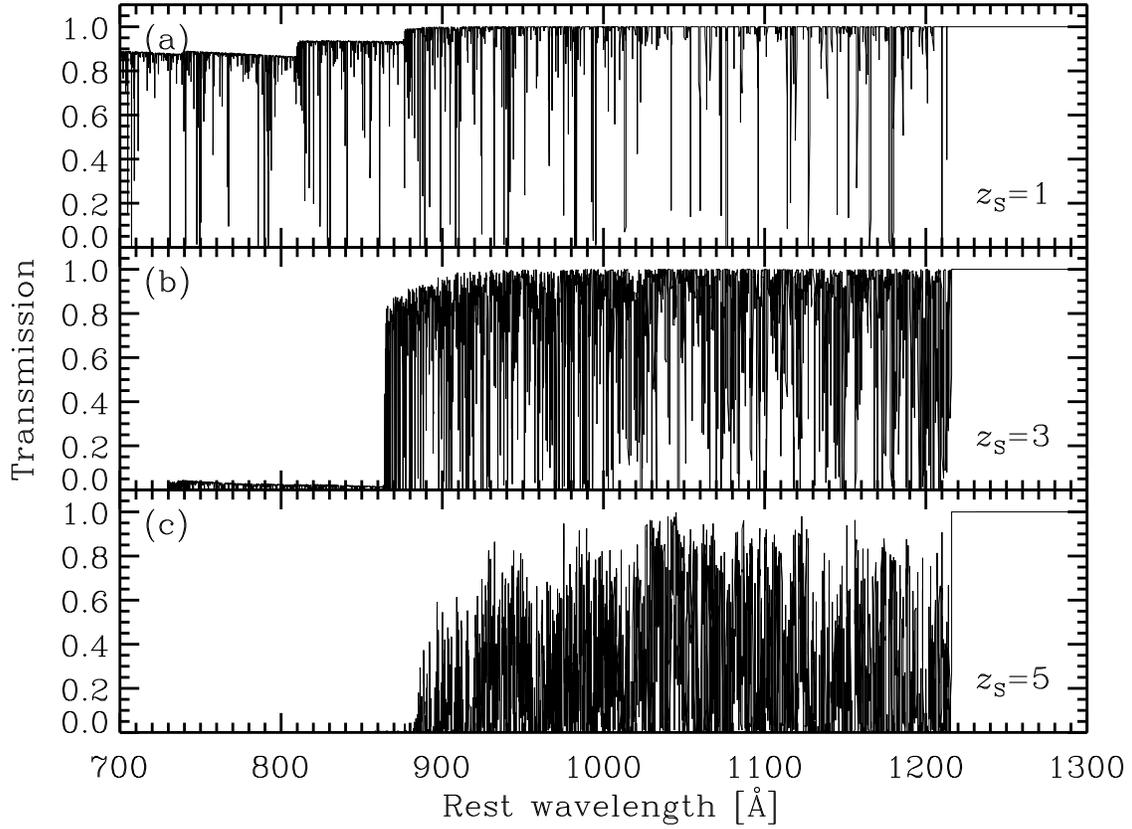}
 \caption{Examples of the IGM transmissions along a line of sight. 
 The horizontal axis is the wavelength in the source rest-frame. 
 The source redshifts are noted in the panels.}
\end{figure*}

\begin{figure*}
 \includegraphics[width=15cm]{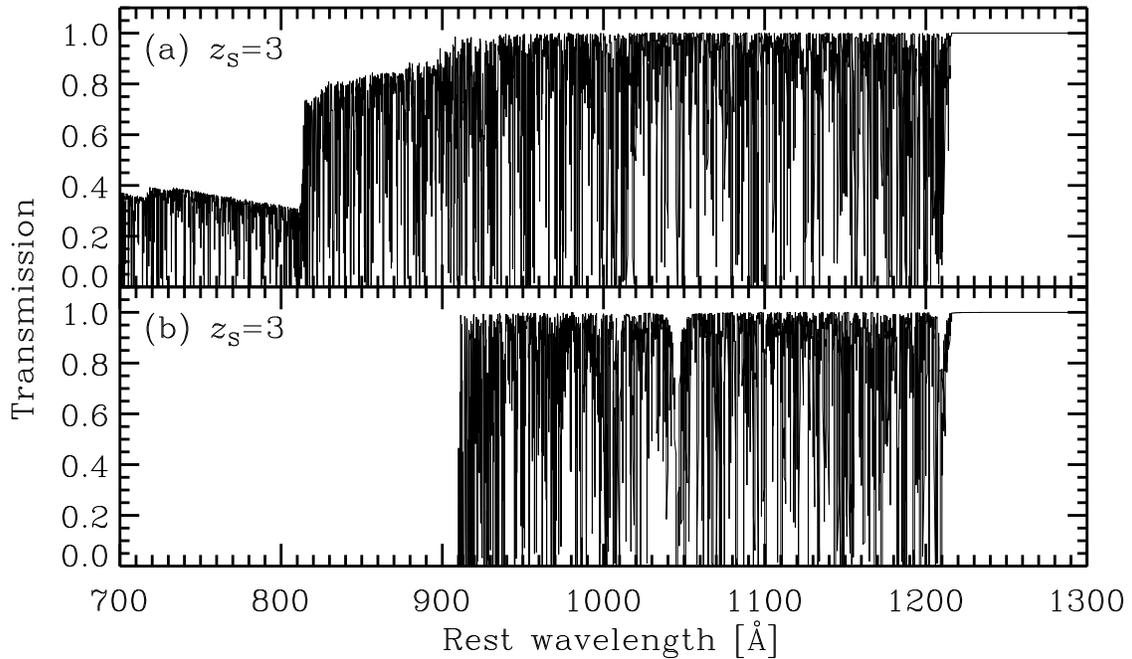}
 \caption{Example IGM transmissions of the source redshift $z_{\rm S}=3$
 for different lines of sight.}
\end{figure*}

\begin{figure*}
 \includegraphics[width=15cm]{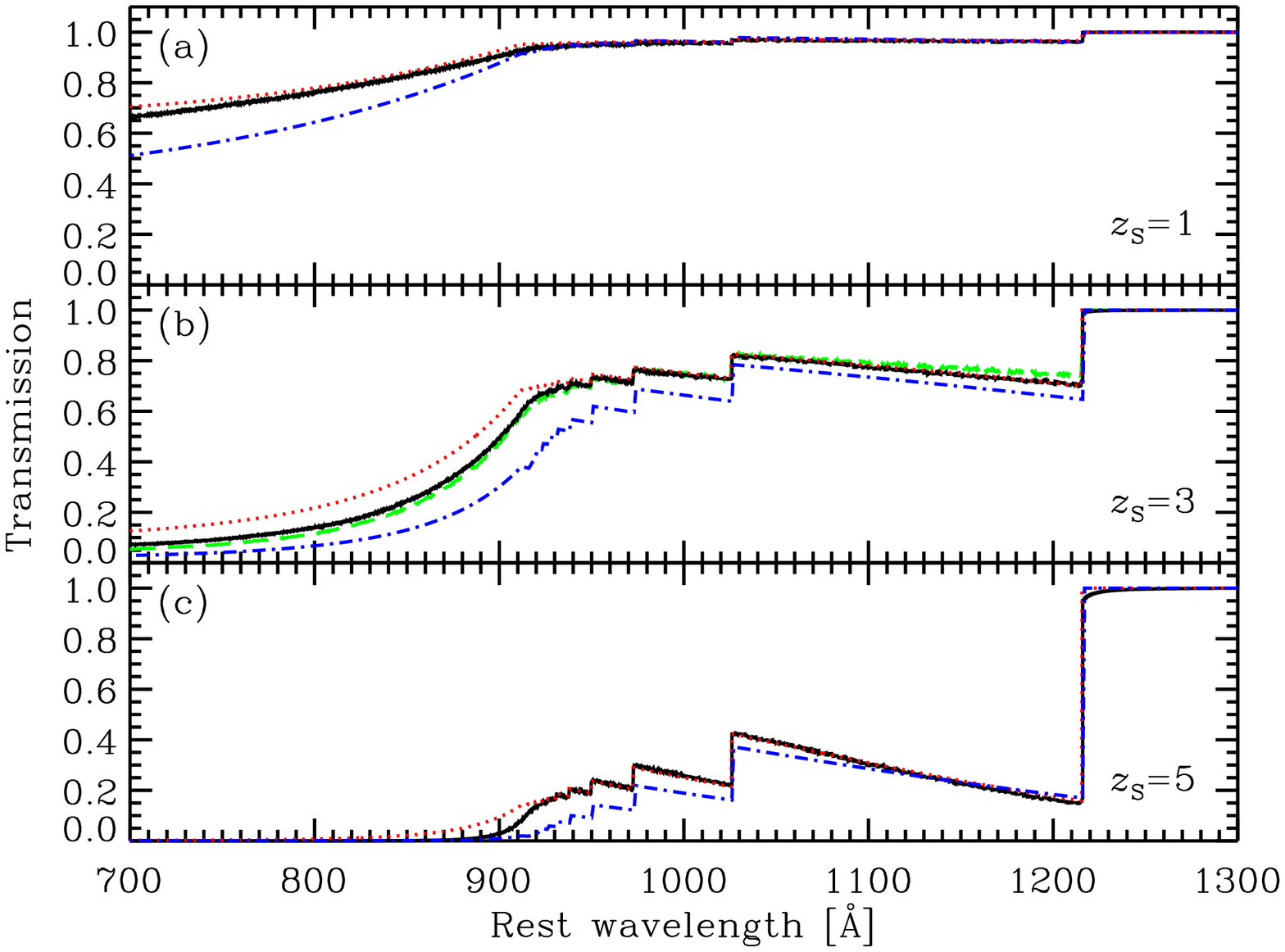}
 \caption{Average IGM transmissions.  The horizontal axis is
 the wavelength in the source rest-frame. The source redshifts are noted
 in the panels. The solid lines are our Monte Carlo results; 10,000
 lines of sight are averaged in each panel. The dot-dashed lines are the
 mean transmission models of Madau (1995). The dotted lines are the mean
 transmission models of Meiksin (2006) but the updated version. The
 dashed line in the panel (c) is the average transmission of the Monte
 Carlo simulation by Bershady et al.~(1999) (MC-Kim model).}
\end{figure*}

On the other hand, such stochastic behaviour disappears in average
transmissions shown in Figure~6, where we also compare our models 
with those by other authors. The solid line in each panel is our Monte
Carlo model. We have averaged 10,000 lines of sight in each panel. Thus,
statistical variations are suppressed in a very small level. 
The dashed line in the panel (b) is the Monte Carlo model by
\cite{ber99} (MC-Kim model). Although the adopted distribution function
of the absorbers is different from ours, we find a good agreement
between \cite{ber99} and us for the source redshift $z_{\rm S}=3$. 
This is caused by a coincidence of the number density of the absorbers
around $z=3$ in our model with that in their model. For example, the
coincidence of the number of LLSs around $z=3$ in our model with that in
their model shown in Figure~3 makes a very good agreement of
both transmissions below the Lyman limit. 
However, we may expect that some differences appear for other redshifts
because of differences of the number density of the absorbers in other
redshifts.

As the dotted lines in Figure~6, we show the mean transmission models by
\cite{mei06}. This is not the original version but the updated one where
the treatment of the Lyman continuum absorption by the LAF was revised
(A.~Meiksin, private communication). The agreement between the updated
\cite{mei06} and us is excellent for all the redshift in the Lyman
series regime. In the Lyman continuum regime, our simulation
shows smaller transmissions than those of \cite{mei06}. 
This is because the number of LLSs in our model is almost
always larger than that in \cite{mei06} (i.e. the regression line by
\citealt{ste95}, see Figure~3). For example, around $z=3$, the number of
LLSs in \cite{mei06} is about two-thirds of ours and that of
\cite{sar89} which is assumed in \cite{ber99}, so that the transmission
of \cite{mei06} is larger than ours and \cite{ber99} at 
$z_{\rm S}=3$ (panel [b]). 

The dot-dashed lines in Figure~6 are the mean transmission models by
\cite{mad95}. We find that \cite{mad95} model almost always shows
smaller transmissions than recent models. \cite{mad95} calculates
the Lyman series regime from the equivalent width distribution of the
LAF and the Lyman continuum regime from the column density distribution
of the LAF and LLSs. \cite{ber99} find that the Lyman series
transmissions based on the equivalent width distribution of 
\cite{mad95} are smaller than those based on the column density
distribution of \cite{mad95}, and that the transmissions increase
further if the Doppler parameter $b$ distribution is taken into account,
while \cite{mad95} assumed a constant $b=35$ km s$^{-1}$. \cite{tep07}
recently reproduces the Lyman $\alpha$ depression $D_{\rm A}$ for
$0<z<6$ by their Monte Carlo simulation with the column density
distribution of \cite{mad95} not the equivalent width distribution, 
and their model is equivalent with that of \cite{ber99} (but MC-NH model
not MC-Kim model shown in Fig.~6). 
The smaller transmissions of \cite{mad95} in the Lyman continuum regime
for $z_{\rm S}=1$ (panel [a]) are probably caused by the larger number
of LLSs of \cite{sar89} than ours (see Fig.~3).

\subsection{Lyman series and Lyman continuum transmissions}

\begin{figure}
 \includegraphics[width=8cm]{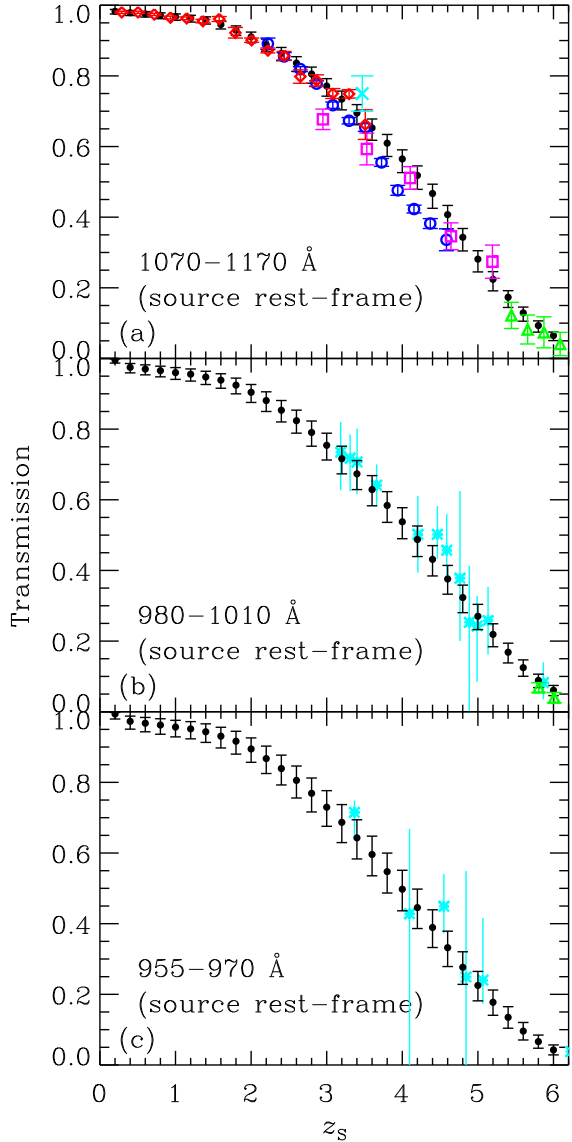}
 \caption{Transmissions averaged over the wavelength ranges blue-ward of
 Ly$\alpha$ (1070--1170 \AA; [a]), Ly$\beta$ (980--1010 \AA; [b]), and
 Ly$\gamma$ (955--970 \AA: [c]) in the source rest-frame. The
 horizontal axis is the source redshift. The filled circles with
 vertical error-bars are median and the central 68\% range of the
 wavelength-averaged transmissions for 10,000 lines of sight generated
 in our Monte Carlo simulation. The open diamonds, circles, triangles,
 and asterisks are taken from Kirkman et al.~(2007), Faucher-Gigu{\`e}re
 et al.~(2008), Fan et al.~(2006), and Songaila (2004),
 respectively. The open squares are the data obtained from
 Tepper-Garc{\'i}a \& Fritze~(2008) but binned and corrected for the
 metal absorptions (the method by Tytler et al.~2004) by us. The cross
 mark is the estimation by us based on the observations by Steidel et
 al.~(2001) (see appendix A).}
\end{figure}

Figure~7 shows transmissions averaged over the wavelength ranges
blue-ward of Lyman $\alpha$, $\beta$, and $\gamma$ lines in the source
rest-frame as a function of the source redshift. The case blue-ward of
Lyman $\beta$ line includes both of Lyman $\alpha$ absorption and Lyman
$\beta$ absorption. The case blue-ward of Lyman $\gamma$ line includes
Lyman $\gamma$ absorption as well as Lyman $\alpha$ and $\beta$
absorptions. Note that the transmissions shown in Figure~7 are
wavelength-averaged ones, but those in Figure~6 are averaged over lines
of sight. The filled circles with error-bars denote the median and the
central 68\% range in the distribution of the wavelength-averaged
transmissions of 10,000 lines of sight for each source redshift. In the 
panels, we also show observational estimations for comparisons. We find
an excellent agreement between our simulation and the observed data
although a dispersion of the observed data is still found at 
$z_{\rm S}\ga3$. This excellent agreement ensures the validity of our
model for the Lyman series absorption. 

There are small upwards shifts of the filled circles (our model
predictions) at the source redshift $z_{\rm S}=0.2$ relative to those at
$z_{\rm S}\ge0.4$ in the panels (b) and (c). These features are  
due to the lack of the Lyman $\alpha$ absorptions in the considering 
wavelength ranges. For example, to absorb the radiation of the Lyman 
$\beta$ range (980--1010 \AA) of $z_{\rm S}=0.2$ by Lyman $\alpha$ line, 
we would need absorbers with negative redshift. Since we consider 
the absorbers only at a positive redshift, the transmission in the
Lyman $\beta$ range shows such a small jump between $z_{\rm S}=0.2$ and
$0.4$. The same is true for the Lyman $\gamma$ transmission.

\begin{figure}
 \includegraphics[width=8cm]{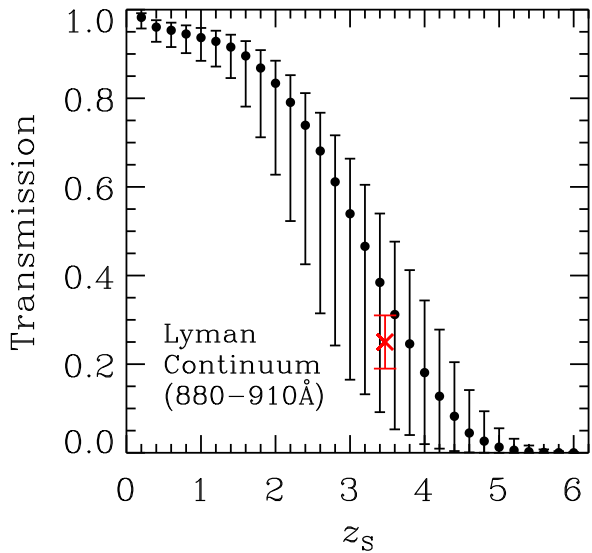}
 \caption{Transmissions averaged over the wavelength range of the Lyman
 continuum (880--910 \AA) in the source rest-frame. The horizontal axis
 is the source redshift. The filled circles and vertical error-bars are
 median and the central 68\% range of the wavelength-averaged
 transmissions for 10,000 lines of sight generated in our Monte Carlo
 simulation. The cross mark is the estimation by us based on the
 composite spectrum of 15 QSOs at $\langle z \rangle=3.47$ by Steidel et
 al.~(2001) (see appendix A).}
\end{figure}

Figure~8 shows transmissions averaged over the wavelength range of the
Lyman continuum in the source rest-frame as a function of the source
redshift. This wavelength-averaged transmissions include all the Lyman
series absorption and the Lyman continuum absorption. The filled circles
and error-bars are the median and the central 68\% range of the
distribution of the wavelength-averaged transmissions of 10,000
lines of sight for each source redshift. We first find that the
dispersion is quite larger than those of Lyman series wavelengths shown
in Figure~7. The stochastic nature of the Lyman continuum absorption
results in such a large variance. We also show the observed estimation
of the Lyman continuum transmission by \cite{ste01} in the figure. This
is based on the composite spectrum of 15 QSOs at 
$\langle z \rangle=3.47$ (see appendix A for details). 
The value is compatible with our Monte Carlo
simulation although it is somewhat smaller than the median of our
simulation. In order to examine the validity of our model for the Lyman
continuum absorption, we need other independent observational
estimations of the Lyman continuum transmissions, which seems quite rare
in the literature at the moment. We would encourage observational
measurements of the Lyman continuum transmissions. The small jump in our
model between the source redshift $z_{\rm S}=0.2$ and $0.4$ because of
the same reason of Figure~7 (b) and (c).

\subsection{Opacity distribution functions}

\begin{figure*}
 \includegraphics[width=16cm]{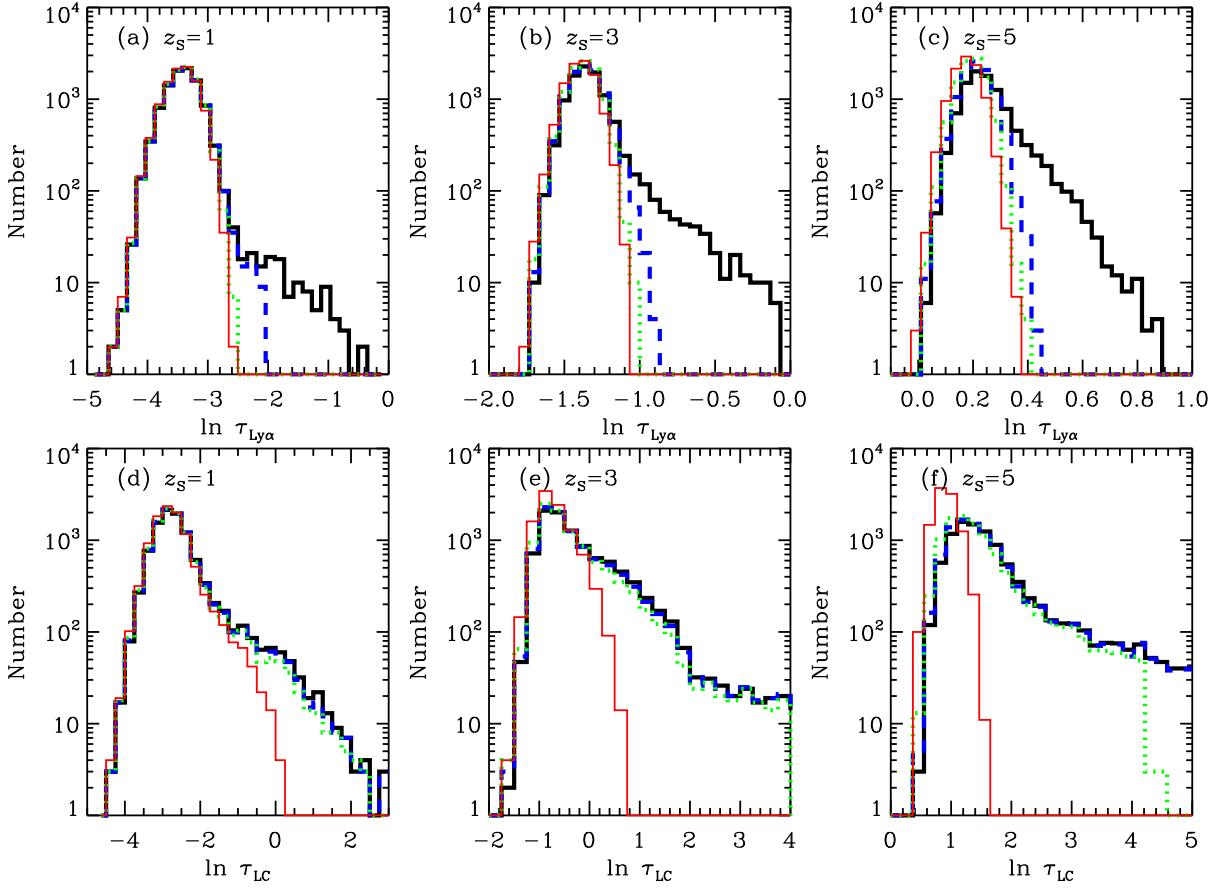}
 \caption{Histograms of the logarithm of effective optical depths in the
 wavelength ranges blue-ward of Ly$\alpha$ (1070--1170 \AA; [a]--[c])
 and of the Lyman limit (880--910 \AA; [d]--[f]) in the source
 rest-frame. The source redshifts are noted in the panels. The thin
 solid histograms are the cases with only the LAF contribution (i.e. 
 $\log_{10} N_{\rm HI}/{\rm cm^{-2}}<17.2$). The dotted histograms are the
 cases with the LAF and LLSs ($\log_{10} N_{\rm HI}/{\rm cm^{-2}}<19.0$). 
 The dashed histograms are the cases without DLAs 
 ($\log_{10} N_{\rm HI}/{\rm cm^{-2}}<20.3$). 
 The thick solid histograms are the cases with all types of absorbers. 
 Each histogram consists of 10,000 lines of sight.}
\end{figure*}

Figure~9 shows histograms of the logarithm of effective optical depths
($\ln \tau_{\rm eff}$) corresponding to the wavelength-averaged
transmission shown in Figures~7 and 8 
(i.e. $\tau_{\rm eff} \equiv -\ln \langle T \rangle$), of the
Lyman $\alpha$ regime (the panels [a]--[c]) and of the Lyman continuum
regime (the panels [d]--[f]) for three source redshifts. 
The thick solid histograms are the cases where all types of absorbers
are taken into account. On the other hand, the dashed histograms are the
cases without DLAs, the dotted histograms are the cases without DLAs and
sub-DLAs, and the thin solid histograms are the cases of only the LAF
contribution. 

The histograms of $\ln \tau_{\rm Ly\alpha}$ of all absorbers (thick
solid ones in panels [a]--[c]) show a broad tail towards large opacity,
whereas the histograms without DLAs do not show such prominent tails.
Thus, we conclude that these tails are mainly caused by DLAs 
($N_{\rm HI}>2\times10^{20}$). Since the number density of DLAs
increases along the redshift as shown in Figure~2, these tails become
more significant for higher redshifts. The $z_{\rm S}=5$ histogram of
all absorbers seems largely modulated by the tail, whereas the fraction
of lines of sight with $\ln \tau_{\rm Ly\alpha}>-1.0$ for $z_{\rm S}=3$
is 5\%, and the fraction of lines of sight with 
$\ln \tau_{\rm Ly\alpha}>-2.5$ for $z_{\rm S}=1$ is 1.4\%. 
We also note that the effect of LLSs and sub-DLAs on $\tau_{\rm Ly\alpha}$ 
is small. This is already found by \cite{tep07} with their Monte Carlo
simulation and explained very well with the curve-of-growth theory. 
Since they did not include DLAs in their simulation, however, they
could not find the tails found here.

If we exclude DLAs, the histograms of $\ln \tau_{\rm Ly\alpha}$ (thin
solid, dotted, and dashed histograms in panels [a]--[c]) seems Gaussian
apparently. Note that the Gaussian distribution becomes parabola in the
panels since the $y$-axis is shown in logarithmic scale. Suppose a null 
hypothesis that the histogram of $\ln \tau_{\rm Ly\alpha}$ is Gaussian
in order to examine the Gaussianity quantitatively.  
Based on the Kolmogorov-Smirnov (K-S) test, we cannot reject the null
hypothesis with a significance level of 1\% for the cases of only the
LAF and of the LAF+LLSs at $z_{\rm S}\ge2$. Thus, the distribution of
$\tau_{\rm Ly\alpha}$ for these cases can be regarded as log-normal. 
For the $z_{\rm S}=1$ cases, however, we can reject the null hypothesis
with a significance level of 0.01\% by the K-S test. 
If we include sub-DLAs, Gaussianity of the distribution of 
$\ln \tau_{\rm Ly\alpha}$ is reduced, but few cases can be regarded as
Gaussian (i.e. we cannot reject the null hypothesis with a significance 
level of 1\% for the cases).

We also find that the dispersion of $\ln \tau_{\rm Ly\alpha}$ decreases 
along the redshift. Note that the displayed range of the $x$-axis becomes
narrower as the redshift becomes larger. Such a log-normal behaviour
with a decreasing variance in $\ln \tau_{\rm Ly\alpha}$ is found in the
observed probability distribution function of the transmitted flux of
the QSOs' spectra \citep{bec07}. On the other hand, the dispersion of
transmissions shown in Figure 7 increases from $z=0$ to $z=3$--4 and
decreases towards higher redshifts (see also \citealt{tep07}). 
This difference between the transmission distribution and the
logarithmic opacity distribution is caused by logarithmic
transformations from transmission to logarithmic opacity.

\cite{tep07} suggest that the distribution of the wavelength-averaged
transmission for higher redshifts approaches Gaussian because
of the Central Limit Theorem; sum of the contributions of a large number
of absorbers. Indeed, the histograms of the wavelength-averaged
transmission in the Lyman $\alpha$ regime of only the LAF case and of
the LAF+LLSs case in our simulation could be regarded as Gaussian; we
cannot reject the Gaussian hypothesis with a significance level of 1\%
for these cases at $z_{\rm S}\ge3$ based on the K-S test. However, we
also find with the K-S test that the distribution of 
$\ln \tau_{\rm Ly\alpha}$ is closer to Gaussian than that of the
wavelength-averaged Lyman $\alpha$ transmission.

The histograms of $\ln \tau_{\rm LC}$ (Fig.~9 [d]--[f]) show 
a broad tail towards large opacity. Unlike $\ln \tau_{\rm Ly\alpha}$,
these tails are mainly produced by LLSs. Even in the LAF only case, we
find such tails. Thus, the histograms of $\ln \tau_{\rm LC}$ are not 
Gaussian. 
These tails make the mode of the opacity always
smaller than the mean. Thus, we have a large probability having a
transmission larger than that expected from the mean opacity model. We
will come back to this point in section 5 where we discuss the
detectability of the Lyman continuum of distant galaxies. 
We note that the cut-off at $\ln \tau_{\rm LC}\approx4$ found in the
LAF+LLSs histograms of panels [e] and [f] corresponds to the optical
depth for the upper limit of HI column density of LLSs (i.e. 
$N_{\rm HI}\le1\times10^{19}$ cm$^{-2}$ and $\tau_{\rm LC}\le63$ for a
single LLS).

\subsection{Estimating Lyman continuum opacity from Lyman $\alpha$
  opacity}

Can we estimate the Lyman continuum opacity from the Lyman $\alpha$
opacity? If it is possible, we may estimate the former opacity for an
individual distant galaxy because we can estimate the latter opacity
from the observed rest-frame 1000--1500 \AA\ spectrum individually.

Figure~10 (a) shows the probability distribution in the plane of the
logarithms of the effective Lyman $\alpha$ opacity ($\tau_{\rm Ly\alpha}$) 
and the effective Lyman continuum opacity ($\tau_{\rm LC}$). This is
produced by 300,000 sets of $(\tau_{\rm Ly\alpha},\tau_{\rm LC})$: 30
source redshifts (0.2--6.0 with 0.2 interval) times 10,000 lines of sight. 
Contrary to the note by \cite{sha06} based on their Monte Carlo
simulation of the intergalactic absorption, we find a good correlation
between $\ln\tau_{\rm Ly\alpha}$ and $\ln\tau_{\rm LC}$; the correlation
coefficient is 0.860. Since \cite{sha06} simulated the
intergalactic absorption only for a single source redshift 
$z_{\rm S}=3.06$, the dynamic range of $\tau_{\rm Ly\alpha}$ might 
be too small for them to find the correlation. Indeed, the probability
density for $z_{\rm S}=3$ shown as the dotted contour in Figure 10 (a)
is confined in a small range of $\tau_{\rm Ly\alpha}$ and elongated
towards large $\tau_{\rm LC}$. Although LLSs mainly
control $\tau_{\rm LC}$ and the LAF does $\tau_{\rm Ly\alpha}$, the LAF
still has an effect on $\tau_{\rm LC}$. Thus, the correlation between 
$\tau_{\rm LC}$ and $\tau_{\rm Ly\alpha}$ seems natural. However, we
should confirm it observationally in future. Currently, we have one
estimate (cross mark in the figure) based on the composite spectrum of
15 QSOs at $\langle z \rangle = 3.47$ by \cite{ste01}, which is
consistent with our prediction.

In Figure~10 (a), we also show the median and the central 68\% range of
the $\tau_{\rm LC}$ distribution for 12 $\tau_{\rm Ly\alpha}$
bins. Table~2 is the summary of them. Interestingly, we may estimate a
statistically probable range of $\tau_{\rm LC}$ and Lyman continuum
transmission from $\tau_{\rm Ly\alpha}$. We should note that the
probable ranges of $\tau_{\rm LC}$ depend on the wavelength range
considered. In Figure~10 (b), we show the difference of the medians and
the 68\% ranges for two cases of the wavelength range in the Lyman
continuum: filled symbols are the case of 880--910 \AA\ and open symbols
are the case of 760--910 \AA. For a wider wavelength case, we expect a
larger optical depth (i.e. smaller transmission) because of a larger
probability having a LLS in the considering wavelength range. Thus, we
should care the wavelength range to estimate $\tau_{\rm LC}$ from 
$\tau_{\rm Ly\alpha}$. In Table~2, we have taken the wavelength range of
880--910 \AA\ in the source rest-frame.

\begin{figure*}
 \includegraphics[width=15cm]{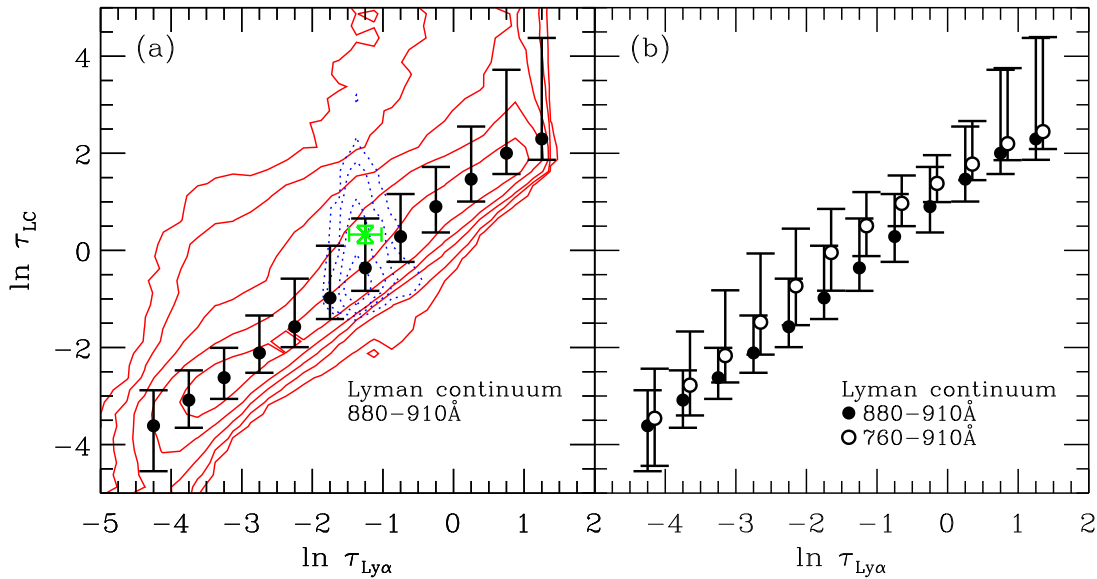}
 \caption{(a) Correlation between the effective optical depths in the
 wavelength ranges blue-ward of Ly$\alpha$ (1070--1170 \AA; horizontal
 axis) and of the Lyman limit (880--910 \AA; vertical axis) in the
 source rest-frame. The contours show the probability density in this
 plane and the contour levels are $0.001$, $0.003$, $0.01$, $0.03$, and
 $0.1$ from the outside; the solid contour is for $z_{\rm S}=0.2$--6 and
 the dotted contour is for $z_{\rm S}=3$. The filled circles and the
 error-bars indicate the position of the median and the width of the
 central 68\% range of the distribution in each bin of the horizontal
 axis (see Table~2). The cross mark is the estimation by us based on the
 QSO composite spectrum by Steidel et al.~(2001) (see appendix A).  
 (b) Difference of the medians and the central 68\% ranges for two
 wavelength ranges of the Lyman continuum; filled symbols are the case of
 880--910 \AA\ and open symbols are the case of 760--910 \AA. The open
 symbols are offset to the right by 0.1 dex for graphical clarity.}
\end{figure*}

\begin{table}
 \caption[]{Median and central 68\% width of the distribution of 
 $\ln \tau_{\rm LC}$ and corresponding Lyman continuum transmission as a
 function of $\ln \tau_{\rm Ly\alpha}$.}
 \setlength{\tabcolsep}{3pt}
 \footnotesize
 \begin{minipage}{\linewidth}
  \begin{tabular}{lcccccc}
   \hline
   & \multicolumn{3}{c}{$\ln \tau_{\rm LC}$} &
   \multicolumn{3}{c}{transmission rate}\\
   $\ln \tau_{\rm Ly\alpha}$ & median & lower & upper 
   & median & lower & upper \\
   \hline
   $-4.25$ & $-3.61$ & $-4.55$ & $-2.88$ & $0.973$ & $0.945$ & $0.989$ \\
   $-3.75$ & $-3.08$ & $-3.65$ & $-2.47$ & $0.955$ & $0.919$ & $0.974$ \\
   $-3.25$ & $-2.62$ & $-3.06$ & $-2.01$ & $0.930$ & $0.875$ & $0.954$ \\
   $-2.75$ & $-2.11$ & $-2.52$ & $-1.34$ & $0.886$ & $0.770$ & $0.923$ \\
   $-2.25$ & $-1.57$ & $-1.99$ & $-0.58$ & $0.812$ & $0.571$ & $0.872$ \\
   $-1.75$ & $-0.98$ & $-1.41$ & ${\phantom -}0.10$ 
   & $0.687$ & $0.331$ & $0.783$ \\
   $-1.25$ & $-0.36$ & $-0.83$ & ${\phantom -}0.66$ 
   & $0.498$ & $0.144$ & $0.647$ \\
   $-0.75$ & ${\phantom -}0.28$ & $-0.24$ & ${\phantom -}1.16$ 
   & $0.266$ & $0.041$ & $0.455$ \\
   $-0.25$ & ${\phantom -}0.90$ & ${\phantom -}0.37$ & ${\phantom -}1.72$ 
   & $0.085$ & $0.003$ & $0.235$ \\
   ${\phantom -}0.25$ & ${\phantom -}1.47$ & ${\phantom -}1.01$ 
   & ${\phantom -}2.55$ & $0.013$ & $0.000$ & $0.064$ \\
   ${\phantom -}0.75$ & ${\phantom -}2.00$ & ${\phantom -}1.57$ 
   & ${\phantom -}3.72$ & $0.001$ & $0.000$ & $0.008$ \\
   ${\phantom -}1.25$ & ${\phantom -}2.30$ & ${\phantom -}1.87$ 
   & ${\phantom -}4.38$ & $0.000$ & $0.000$ & $0.002$ \\
   \hline
  \end{tabular}

  \medskip

  The wavelength ranges are 1070--1170 \AA\ for Ly$\alpha$ and 
  880--910 \AA\ for the Lyman continuum in the source rest-frame.

 \end{minipage}
\end{table}%

\section{Detectability of Lyman continuum}

Although the observations of the Lyman continuum is very challenging
because of the intergalactic absorption as well as the interstellar one,
\cite{sha06} did detect the continuum from 2 out of 14 Lyman
break galaxies at $z\sim3$. On the other hand, \cite{mal03} and
\cite{sia07} did not detect the continuum from total 32 
star-forming galaxies at $z\sim1$. Combining these results of direct
observations with an indirect estimation from the ionising background
intensity, \cite{ino06} suggested an evolution of the cosmic
average escape fraction along the redshift: larger escape fraction at 
higher redshift on average. Such an evolution scenario of the average
escape fraction seems to be supported by a recent cosmological simulation
\citep{raz06,raz07}. \cite{sha06} showed that some galaxies do have a
large escape fraction. If the number of the galaxies with a large escape
fraction increases along the redshift, such galaxies may contribute to
the cosmic reionisation significantly. Thus, understanding the
properties of such galaxies is very interesting. However, any common
property in two detected galaxies of \cite{sha06} were not found. We may
need much more number of galaxies showing a large escape of the Lyman
continuum. Here, we discuss the detectability of such galaxies at
$z\ge1$ using the intergalactic transmission presented in the previous
section. 

The observable flux density of the Lyman continuum is expressed as 
\begin{equation}
 F_{\rm LC}^{\rm obs} = {\cal R}_{\rm esc} F_{\rm UV}^{\rm obs} 
  T_{\rm LC}^{\rm IGM}\,,
\end{equation}
where ${\cal R}_{\rm esc}$ is the escaping flux density ratio of the
Lyman continuum to the non-ionising ultraviolet introduced by
\cite{ino06}, $F_{\rm UV}^{\rm obs}$ is the observed non-ionising
ultraviolet flux density, and $T_{\rm LC}^{\rm IGM}$ is the
intergalactic transmission of the Lyman continuum. We may use the escape
ratio ${\cal R}_{\rm esc}$ as a proxy of the absolute escape fraction
because they are the same order usually \citep{ino06}. 
Estimating the absolute escape fraction is difficult in general 
because we need the dust attenuation correction and the intrinsic 
spectrum, while ${\cal R}_{\rm esc}$ is just the observed flux density 
ratio corrected for the IGM opacity (see \citealt{ino06} for details). 

Suppose an observation in the Lyman continuum of distant galaxies with a
limiting flux density $F_{\rm LC}^{\rm lim}$. In other words, we can
detect a galaxy with its flux density in the Lyman continuum of 
$F_{\rm LC}^{\rm obs} \ge F_{\rm LC}^{\rm lim}$ with an enough
significance. In this case, the IGM transmission towards the galaxy must
satisfy 
\begin{equation}
 T_{\rm LC}^{\rm IGM} \ge T_{\rm LC}^{\rm lim}
  \equiv \frac{F_{\rm LC}^{\rm lim}}
  {{\cal R}_{\rm esc} F_{\rm UV}^{\rm obs}}\,.
\end{equation}
The quantity $T_{\rm LC}^{\rm lim}$ means the minimum IGM transmission
for a galaxy to be detected by an observation. Thus, the detectable
probability of a galaxy is the probability that we have a transmission
equal to or larger than $T_{\rm LC}^{\rm lim}$. The detectable
probability becomes larger for a fainter $F_{\rm LC}^{\rm lim}$
(i.e. deeper observation), a larger ${\cal R}_{\rm esc}$ (i.e. larger
escape fraction), or a brighter $F_{\rm UV}^{\rm obs}$ (i.e. more
luminous object).

Figure~11 shows the probability of lines of sight with a Lyman continuum 
transmission $T_{\rm LC}^{\rm IGM}$ larger than the value indicated in
the horizontal axis. This is just a cumulative probability calculated
from 10,000 transmissions produced by our Monte Carlo simulation for
each source redshift $z_{\rm S}$. We show 5 cases of $z_{\rm S}=1$--5 as
the thick solid lines. The case of $z_{\rm S}=6$ is always zero
probability in the range of the horizontal axis shown in the figure. We
expect that the fraction of lines of sight not opaque against the Lyman
continuum from $z_{\rm S}=3$ (i.e. $T_{\rm LC}^{\rm IGM}>0.37$ for 
$\tau_{\rm LC}<1$) is 70\%, and the fraction even for $z_{\rm S}=3.8$ is
25\%.

\begin{figure}
 \includegraphics[width=8cm]{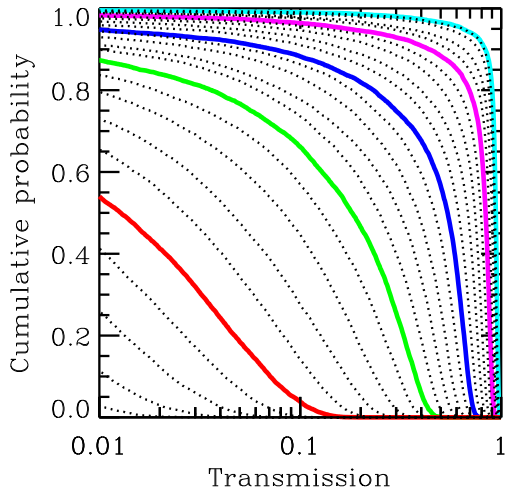}
 \caption{Probability of having a line of sight with a transmission
 averaged over the Lyman continuum (880--910 \AA) in the source rest-frame
 larger than the value indicated in the horizontal axis. The lines from
 right to left are the cases of the source redshift from 1.0 to 5.8 with
 the redshift interval of 0.2. The cases of the source redshift of 1, 2,
 3, 4, and 5 are indicated as the thick solid lines.}
\end{figure}

\begin{table}
 \caption[]{Detectable probability of a galaxy with a large escape
 fraction.}
 \setlength{\tabcolsep}{3pt}
 \footnotesize
 \begin{minipage}{\linewidth}
  \begin{tabular}{lccccccc}
   \hline
   & \multicolumn{7}{c}{$m_{\rm LC}^{\rm limit}-m_{\rm UV}^{\rm obs}$ (AB)}
   \\
   $z_{\rm S}$ & 1.5 & 2.0 & 2.5 & 3.0 & 3.5 & 4.0 & 4.5 \\
   \hline
   1.0 & 90 & 96 & 98 & 99 & 99 & 99 & 99 \\
   2.0 & 49 & 88 & 92 & 94 & 96 & 97 & 97 \\
   3.0 & 0 & 52 & 73 & 81 & 86 & 89 & 91 \\
   4.0 & 0 & 0 & 19 & 45 & 59 & 70 & 76 \\
   5.0 & 0 & 0 & 0 & 0 & 1 & 7 & 17 \\
   \hline
  \end{tabular}

  \medskip

  The detectable probability (\%) of a galaxy with a large escape
  ratio of ${\cal R}_{\rm esc}=0.3$ at the redshift $z_{\rm S}=1$--5 
  when observing the wavelength range of 880--910 \AA\ in the source
  rest-frame. Seven cases of the limiting magnitude of the Lyman
  continuum observation are shown: 
  $m_{\rm LC}^{\rm limit} = m_{\rm UV}^{\rm obs}+1.5$, 2.0, 2.5,
  3.0, 3.5, 4.0, and 4.5, where $m_{\rm UV}^{\rm obs}$ is the observed
  magnitude in the non-ionising ultraviolet ($\sim$1500 \AA\ in the
  rest-frame) of the galaxy. We expect a larger probability for a deeper
  observation (i.e. larger $m_{\rm LC}^{\rm limit}$) or a more luminous
  object (i.e. smaller $m_{\rm UV}^{\rm obs}$). Note that the detectable
  probability depends on the observing wavelength range (see text).
 \end{minipage}
\end{table}%

Table~3 is a summary of the expected detectable probability (\%) with
the assumption of ${\cal R}_{\rm esc}=0.3$ which is estimated from the
two detections of \cite{sha06}. When observing the Lyman continuum of
880--910 \AA\ in the source rest-frame, we find that 70--80\% galaxies
with a large escape fraction of ${\cal R}_{\rm esc}=0.3$ are detectable
for $z_{\rm S}=3$ and 20--45\% galaxies are detectable even for
$z_{\rm S}=4$ if the objects are 24.5 AB in UV and the limiting
magnitude of the Lyman continuum is 27.0--27.5 AB. In the observations
of \cite{sha06}, the limiting magnitude of the Lyman continuum and the
average observed UV magnitude are 27.56 AB and 23.92 AB, respectively,
and the average redshift is 3.06; the detectable probability of a galaxy
with ${\cal R}_{\rm esc}=0.3$ in their sample is about 90\%. Since
\cite{sha06} detected 2 out of 14 galaxies, therefore, we expect that
the fraction of the galaxies with ${\cal R}_{\rm esc}=0.3$ is about 16\%
in their sample. This suggests that there is a large variance of the
escape fraction at $z\sim3$, which is very interesting and whose origin
should be understood in the context of the galaxy evolution in future.

We note that the detectable probability in Table~3 may be the most
optimistic case because the probability depends on the observing
wavelength range. If we observe a wider wavelength range, the
probability of having a LLS in the corresponding redshift range
increases, then, the transmission and the detectable probability
decreases. In addition, if we observe a wavelength range far from the
Lyman limit (e.g., $\sim700$ \AA), the probability affected by a LLS
also increases, then, the detectable probability decreases (see also
Figs.~4-6 and 10). Therefore, we should take into account the observing
wavelength range properly when estimating the detectable probability for
future observations.

\section{Conclusions}

Motivated by recent attempts for determining the escape fraction of the
 Lyman continuum from distant galaxies, we have made a Monte Carlo
 simulation of the intergalactic absorption in order to model the Lyman
 continuum absorption properly.
For this simulation, we have derived an empirical distribution function
 of the intergalactic absorbers, Lyman $\alpha$ forest (LAF), Lyman
 limit systems (LLSs), and damped Lyman $\alpha$ systems (DLAs), from
 the observed statistics of the absorbers.
In the distribution function of the absorbers, we have assumed a single
 functional form of the redshift distribution for all types of the
 absorbers, while the previous models of the intergalactic absorption
 assumed two different functional forms for the LAF and LLSs (and DLA).  

The transmission functions obtained from our simulation are very
consistent with the previous models except for \cite{mad95} which
predicts smaller transmissions than others. 
The Lyman series transmissions by our simulation excellently agree
with the observational data, which ensures validity of our model.
We have predicted the Lyman continuum transmissions as a function
of the source redshift. Although observational estimates of the Lyman
continuum transmission are quite rare in the literature, one
observational estimate based on \cite{ste01} is consistent with our
prediction. 

The distribution function of the Lyman $\alpha$ opacity for a source
redshift seems log-normal with a tail towards large opacities. This tail
is produced by DLAs. Since the number density of DLAs increases along
the redshift, the significance of the tail in the distribution function
increases. As found by \cite{tep07}, we also find that the effect of
LLSs on the Lyman $\alpha$ opacity is small. Along the source redshift,
the mean of the log-normal part increases but the variance
decreases. These features are found in recent observations. 

The distribution function of the Lyman continuum opacity shows a very
broad tail towards a large opacity which is produced by LLSs. Unlike the
Lyman $\alpha$ opacity, the effect of DLAs on the Lyman continuum
opacity is small. Rarity of LLSs controlling the Lyman continuum opacity
provides us with a chance to have a clean line of sight for $z\sim4$; 
the probability of a clean (i.e. optical depth less than unity)
line of sight at 900 \AA\ in the source rest-frame is about 70\% for the
source redshift $z\sim3$ and about 20\% for $z\sim4$.

A good correlation between the Lyman $\alpha$ opacity and the Lyman
continuum one is found although \cite{sha06} noted no correlation in
their simulation. This may be because a small dynamic range of the Lyman
$\alpha$ opacity in their simulation. Based on the correlation we find, 
we may predict a statistically probable range of the Lyman continuum
opacity as a function of the Lyman $\alpha$ opacity.  
This may be useful to estimate the Lyman continuum opacity from the
Lyman $\alpha$ one for individual galaxy.

Finally, we have predicted the detectable probability of a galaxy with a
large escape fraction of the Lyman continuum based on our simulation. 
For example, the detectable probability is 86\% when the limiting
magnitude of the Lyman continuum observations is 3.5 AB deeper than the
non-ionising ultraviolet magnitude of the sample galaxy with an escape
ratio ${\cal R}_{\rm esc}=0.3$ at $z=3$. The small number fraction of
the galaxies detected in the Lyman continuum by \cite{sha06} (2 out of
14 sample galaxies) suggests a large variance of the escape fraction at
$z=3$. This should be understood in the context of the galaxy evolution
in future.

\section*{Acknowledgments}

We would like to thank J.-M.~Deharveng for stimulating discussions,  
C.~P{\'e}roux for some comments on the distribution function of the
intergalactic absorbers, A.~Meiksin for giving his updated absorption
model and discussing several points on modelling, M.~Matsumoto for
distributing C-codes of the ``Mersenne Twister'' on his web page, and
the referee, T.~Tepper-Garc{\'i}a for interesting discussions and useful
comments. AKI and II are supported by Grand-in-Aid for Young Scientists
(B).

\appendix

\section{Lyman $\alpha$ and Lyman continuum transmissions based on QSO
 composite spectrum by Steidel et al.~(2001)}

We describe the method for estimating the transmissions in the Lyman
$\alpha$ and the Lyman continuum ranges based on the composite spectrum
of QSOs by \cite{ste01}. The method is essentially same as that used in
\cite{ste01}. However, we assume a different intrinsic spectrum of QSOs
based on the recent observations. The obtained transmissions are plotted
in Figures 7, 8, and 10.

The wavelength ranges of the Lyman $\alpha$ and the Lyman continuum
transmissions in Figures 7, 8, and 10 are 1070--1170 \AA\ and 
880--910 \AA, respectively, in the source rest-frame. We take 1100 \AA\
and 900 \AA\ as their representative values. In general, the IGM 
transmission at a wavelength $x$ is defined as 
$T^{\rm IGM}_x \equiv F^{\rm obs}_x/F^{\rm int}_x$, 
where $F^{\rm obs}_x$ and $F^{\rm int}_x$ are the observed and the
intrinsic flux densities, respectively. The intrinsic flux density can
be estimated from the observed flux density at a wavelength free from
the IGM absorption, that is, $>1216$ \AA, if we assume an intrinsic
spectrum. When we assume the reference wavelength of 1500 \AA, the Lyman
$\alpha$ and the Lyman continuum transmissions are 
\begin{equation}
 T^{\rm IGM}_{\rm Ly\alpha} = 
  \frac{(F_{1100}/F_{1500})_{\rm obs}}{(F_{1100}/F_{1500})_{\rm int}}\,,
\end{equation}
and
\begin{equation}
 T^{\rm IGM}_{\rm LC} = 
  \frac{(F_{900}/F_{1500})_{\rm obs}}{(F_{900}/F_{1500})_{\rm int}}\,.
\end{equation}

We take the observed flux densities from the composite spectrum of QSOs
by \cite{ste01}. The sample consists of 15 QSOs at 
$\langle z \rangle=3.47\pm0.14$. The QSO composite spectrum is
constructed in order to compare the composite spectrum of LBGs from
which \cite{ste01} detected the Lyman continuum of the sample LBGs. 
From Table~2 in \cite{ste01}, we have $F_{1100}^{\rm obs}=1.86\pm0.02$ 
and $F_{900}^{\rm obs}=0.50\pm0.03$ in the arbitrary unit of the flux
density (per Hz). We obtain $F_{1500}^{\rm obs}=3.4$ from their value
of $(F_{1500}/F_{900})_{\rm obs}=6.8\pm0.4$.

The intrinsic spectrum of QSOs is the largest source of the error in
this estimation. The spectra of QSOs can be fit by a composite of
power-laws. We here discuss the indices under the power-law assumption: 
$F_\nu \propto \nu^{\alpha}$. The index in the extreme-ultraviolet
(including the Lyman continuum) is still controversial. \cite{tel02}
obtained $\alpha(500-1200~{\rm \AA})=-1.76\pm0.12$ from 184 QSOs at
$z\sim1$ \citep[see also][]{zhe97}. On the other hand, \cite{sco04}
obtained $\alpha(630-1155~{\rm \AA})=-0.56^{+0.38}_{-0.28}$ from 85 AGNs
at $z\sim0.2$ \citep[see also][]{sha05}. This contrast may be due to an
evolution of the spectrum along the redshift and the luminosity
\citep{sco04}, and also there may be the difficulty of the correction
for the IGM absorption. We here assume two possible indices of 
$\alpha(<1100~{\rm \AA})=-1.8\pm0.1$ and $-0.6\pm0.3$. For longer
wavelength, \cite{zhe97} obtained 
$\alpha(1050-2200~{\rm \AA})=-0.99\pm0.05$. We here adopt 
$\alpha(>1100~{\rm \AA})=-1.0\pm0.2$, taking into account a large
variance of the indices among QSOs \citep[e.g.,][]{sha05}. Therefore, 
the intrinsic ratios of the flux densities adopted here are 
$(F_{1100}/F_{1500})_{\rm int}=0.73\pm0.05$, and 
$(F_{900}/F_{1500})_{\rm int}=0.54\pm0.04$ or $0.69\pm0.11$.

The observed ratios of the flux densities measured in the composite
spectrum of QSOs at $\langle z \rangle = 3.47$ by \cite{ste01} and the
intrinsic ratios described above result in 
$T^{\rm IGM}_{\rm Ly\alpha}=0.75\pm0.05$, and 
$T^{\rm IGM}_{\rm LC}=0.28\pm0.03$ or $0.22\pm0.04$. 
Finally, we summarise the Lyman continuum transmission as 
$T^{\rm IGM}_{\rm LC}=0.25\pm0.06$, where we put the sum of the internal
and external errors as an error estimate.

We should note one point here. In the procedure for obtaining the
intrinsic spectrum of QSOs, we need the IGM absorption correction. Then,
we used the intrinsic spectrum to obtain $T^{\rm IGM}$. Thus, it is like
a tautology although the intrinsic spectrum and $T^{\rm IGM}$ are
obtained from the different data and the different authors.
However, we may consider that the obtained $T^{\rm IGM}$ still have a
meaning since we have assumed a wide range of the spectral indices for
the intrinsic spectrum. For future measurements of $T^{\rm IGM}$, 
we may use an independent intrinsic spectrum of QSOs, for example,
based on an accretion disk model \citep[e.g.,][]{kaw01}.

\label{lastpage}

\end{document}